\documentclass[epsfig,12pt]{article}
\usepackage{epsfig,amssymb,euscript,bbold}
\usepackage{amsmath}

\allowdisplaybreaks
\def\be{\begin{equation}}
\def\ee{\end{equation}}
\def\bea{\begin{eqnarray}}
\def\eea{\end{eqnarray}}

\numberwithin{equation}{section}

\begin{document}
\title{
{\baselineskip -.2in
\vbox{\small\hskip 5in \hbox{hep-th/0610002}}
\vbox{\small\hskip 5in \hbox{DAMTP-2006-74}}
\vskip .01in
\vbox{\small\hskip 5in \hbox{NSF-KITP-06-83}}
\vbox{\small\hskip 5in \hbox{}}} 
\vskip .4in 
At the horizon of a supersymmetric $AdS_5$ black hole: \\
Isometries and half-BPS giants
}
\author{Aninda Sinha $^1$, Julian Sonner $^1$
and Nemani V. Suryanarayana $^2$\\
{}\\
{\small{\it $^1$ Department of Applied Mathematics and Theoretical
    Physics,}}\\ 
{\small{\it Wilberforce Road, Cambridge CB3 0WA, U.K.}} \\
{\small{E-mail: {\tt A.Sinha@damtp.cam.ac.uk}, {\tt
      J.Sonner@damtp.cam.ac.uk}}} \\ 
{\small{$^2$ {\it Perimeter Institute for Theoretical
      Physics,}}}\\ 
{\small{\it 31 Caroline Street North, Waterloo, ON, N2L 2Y5,
    Canada}}\\ 
{\small{E-mail: {\tt vnemani@perimeterinstitute.ca}}}
}
\maketitle
{\vskip 2cm\abstract{\noindent The near-horizon geometry of an
asymptotically $AdS_5$ supersymmetric black hole discovered by
Gutowski and Reall is analysed.  After lifting the solution to 10 dimensions, we explicitly solve
the Killing spinor equations in both Poincar\'e and global
coordinates. It is found that exactly four supersymmetries are preserved which is twice the number for the full black hole. The full set of isometries is constructed and the
isometry supergroup is shown to be $SU(1,1|1) \times SU(2) \times
U(3)$. We further study half-BPS configurations of D3-branes in the near-horizon geometry in  Poincar\'e and global coordinates. Both giant
graviton probes and dual giant graviton probes are found.}}

\newpage
\tableofcontents
%
%
%
%
%
\section{Introduction}
Asymptotically $AdS_5$, rotating, electrically charged supersymmetric
black holes of minimal $D=5$ gauged supergravity with regular horizons
were first constructed by Gutowski and Reall in \cite{gr1,gr2}. These
solutions have  been further generalised in \cite{cclp1, cclp2, klr,hed}. When
lifted to 10-dimensional solutions of type IIB, these geometries
asymptote to the maximally supersymmetric $AdS_5 \times S^5$ solution and preserve just two of the 32 supersymmetries
\cite{ggs}. One of the important outstanding problems in string theory
is to account for the entropy of these black holes both from the
string theory and the holographic boundary gauge theory points of
view.

The standard way of counting the microstates of a supersymmetric black hole in string
theory is to count the BPS states of the D-brane system in the
asymptotic geometry of the black hole. In
recent times it has been realised that the entropy of extremal black
holes depends just on the string theory in the near
horizon geometry called the attractor geometry of the black
hole. Therefore, the Bekenstein-Hawking-Wald entropy should also be related to a certain number of
appropriate BPS states in the attractor geometry or those in the
holographically dual quantum mechanics. This program has
been demonstrated successfully in the context of 4-dimensional
extremal black holes in \cite{gssy, gsy} (see also \cite{kr} and
\cite{ls}). For the supersymmetric $AdS_5$ black holes of \cite{gr1,gr2} , one expects two
complementary approaches to count the microstates as well: (1) count the
BPS states with the right charges in the $AdS_5 \times S^5$
background, (2) count an appropriate set of BPS states in the
near-horizon geometry. The problem of counting BPS states with just
two supersymmetries in $AdS_5 \times S^5$ geometry is a  hard
problem (see \cite{kmmr, bglm, ms} where the problem of counting the
BPS states with four supersymmetries was addressed and \cite{brs} where a fermi-surface model was proposed to achieve qualitative agreement with the counting). In this paper, we
initiate addressing the problem using the second approach.

We consider the single-parameter black holes with equal angular
momenta \cite{gr1} in $AdS_5$ directions and a single $U(1)$ electric
charge.
When the angular momentum vanishes, this solution reduces to $AdS_5$.
We lift the near-horizon geometry of the black hole to a solution of
type IIB supergravity in ten dimensions. By studying the integrability
condition of the Killing spinor equations, it is found that the number
of supersymmetries of the near-horizon geometry is four, which is twice
the number of supersymmetries of the full black hole. We explicitly
construct the Killing spinor in both Poincar\'e and global
coordinates. Using the Killing spinor solution and the technique of
\cite{gmt1, gmt2}, we show that the superisometry group is
$SU(1,1|1)$. We demonstrate that the full isometry supergroup of the
10-dimensional near-horizon solution is $SU(1,1|1)\times SU(2)\times
U(3)$. As one expects for an extremal black hole there is an $AdS_2$
factor in the near-horizon geometry and we consider both Poincar\'e and
global coordinates for it.

We then initiate the study of probe branes in the near-horizon
geometry along the lines of \cite{ls} in the context of BMPV back
holes \cite{bmpv}. Two sets of probe D3 branes are found which
preserve half of the near-horizon supersymmetries. These are the
analogues of giant gravitons \cite{mst} and dual giant gravitons
\cite{dgg1, dgg2} in $AdS_5 \times S^5$. The probes in Poincar\'e
coordinates are static and have vanishing Hamiltonians. They still
carry non-zero angular momenta because of the rotation of the
background. The probes in global coordinates rotate and have non-zero
angular momenta and Hamiltonians. 

The paper is organised as follows. In section 2, we lift the near
horizon solution to ten dimensions. In section 3, we solve the Killing
spinor equation explicitly in Poincar\'e coordinates. In section 4, we
derive the isometry supergroup of the geometry. In section 5, we
consider the problem from the point of view of global coordinates and
solve the Killing spinor equation in these coordinates. In section 6,
we initiate the study of probe branes in Poincar\'e coordinates while
in section 7, the probe branes are studied in global coordinates. We
conclude with a brief discussion in section 8. 

\section{The black hole and its near-horizon geometry}
The metric of the five-dimensional solution with equal angular momenta
is specified by the f\"unfbein \cite{gr1}
\begin{eqnarray}
\label{eqn:fivebhviel}
e^0 &=& {\cal{F}} \big( dt + \Psi \sigma_3^L \big)\,, \quad
e^1 = {\cal{F}}^{-1} (1+{2 \omega^2 \over l^2}+
\frac{r^2}{l^2})^{- \frac{1}{2}} dr \,,\cr
e^2 &=& {r \over 2} \sigma_1^L \,,\quad
e^3 = {r \over 2} \sigma_2^L \,,\quad
e^4 = {r \over 2 l} \sqrt{l^2+2 \omega^2+r^2} \sigma_3^L\,.
\end{eqnarray}
The right-invariant one-forms on SU$(2)$ are $\sigma_1^L = \sin\phi \,
d\theta - \sin\theta \, \cos\phi \, d\psi$, $\sigma_2^L = \cos\phi \,
d\theta + \sin\theta \, \sin\phi \, d\psi$ and $\sigma_3^L = d\phi +
\cos\theta \, d\psi$, where $0 \leq \theta \leq \pi, 0 \leq \psi \leq
2 \pi, 0 \leq \phi \leq 4 \pi$. They satisfy $d\sigma_i^L = -
\frac{1}{2} \epsilon_{ijk} \sigma_j^L \wedge \sigma_k^L$ with
$\epsilon_{123} = 1$. Furthermore
\begin{equation}
{\cal{F}} = 1-{\omega^2 \over r^2}\,,\qquad
%
%
\Psi = -{\eta r^2 \over 2 l}\big( 1+ {2 \omega^2 \over r^2} +{3
  \omega^4 \over 2r^2(r^2-\omega^2)}
\big) \,,
\end{equation}
with $\eta= \pm 1$ and $\omega$ is constant. The 1-form gauge
potential is given by 
\begin{equation}
A = {\sqrt{3} \over 2} \big[ {\cal{F}} dt
+{\eta \omega^4 \over 4 l r^2} \sigma_3^L \big] \,.
\end{equation}
We choose $\eta = 1$ from here on. This solution asymptotes to global
$AdS_5$ and in this limit reads 
\begin{eqnarray}
e^0 = dt - \frac{r^2}{2l} \sigma_3^L, ~~~ e^1 =
\frac{dr}{\sqrt{1+\frac{r^2}{l^2}}}, ~~~ e^2 = \frac{r}{2} \sigma_1^L,
~~~ e^3 = \frac{r}{2} \sigma_2^L, ~~~ e^4 = \frac{r}{2}
\sqrt{1+\frac{r^2}{l^2}} ~ \sigma_3^L
\end{eqnarray}
with $F = dA = 0$. This can be put into the standard form by writing
$\tilde \phi = \phi + \frac{2t}{l}$ and $\tilde t = t$ which imply
$\frac{\partial}{\partial t} = \frac{\partial}{\partial \tilde t} +
\frac{2}{l} \frac{\partial}{\partial \tilde \phi}$. The black hole
solution carries an electric charge under the $U(1)$ gauge field given
by 
\begin{equation}
Q = \frac{1}{4\pi G} \int_{S^3_\infty} \star F = \frac{\sqrt{3} \pi
  \omega^2}{2 G} (1 + \frac{\omega^2}{2l^2}).
\end{equation}
where $G$ is the 5-dimensional Newton's constant. The black hole
carries an angular momentum given by
\begin{equation}
J = \frac{3 \pi \omega^4}{8 l G} (1 + \frac{2\omega^2}{3l^2})\,,
\end{equation}
%
while the entropy is
\begin{equation}
S_{BH} = \frac{\pi}{2G}\omega^3 \sqrt{1 + \frac{3\omega^2}{4l^2}}\,,
\end{equation}
which may be written as \cite{kl}
\begin{equation}
S_{BH} = \sqrt{l^2 Q^2 - \frac{2\pi l^3}{G} \, |J|} = \sqrt{l^2 Q^2 -
   4N^2 \, |J|}\,,  
\end{equation}
in terms of the electric charge and angular momentum of the black hole.
Here $N^2 = \frac{\pi l^3}{2G}$. The near-horizon limit of this geometry is
\begin{eqnarray}\label{eq:nhgframe}
e^0 &=& \frac{2r}{\omega} dt - \frac{3 \omega^2}{4l} \sigma_3^L, ~~~~ 
e^1 = \frac{\omega l}{2 \lambda} \frac{dr}{r}, ~~~~
e^2 = \frac{\omega}{2} \sigma_1^L, ~~~ e^3 = \frac{\omega}{2}
\sigma_2^L, \cr
e^4 &=& \frac{\omega}{2l} \lambda
~ \sigma_3^L, ~~~~  
A = \frac{\sqrt{3}}{2} ( \frac{2r}{\omega} dt + \frac{\omega^2}{4l}
\sigma_3^L) = \frac{\sqrt{3}}{2} (e^0 + \frac{2\omega}{\lambda} e^4)\,.  
\end{eqnarray}
Here we have defined
\be
\lambda=\sqrt{l^2+3\omega^2}\,.
\ee
%
%
The gauge field strength, $F=d A$ is given by
\begin{eqnarray}
F &=& \frac{\sqrt{3}}{2l} [ 3 e^{14} - e^{23} - \frac{2}{\omega}
    \lambda ~ e^{01}], \cr
\star F &=& \frac{\sqrt{3}}{2l} [ 3 e^{023} - e^{014} +
    \frac{2}{\omega} \lambda ~ e^{234}].
\end{eqnarray}
The equations of motion are
\begin{eqnarray}
R_{ab} -2 F_{ac} F_b^{~c} + \frac{1}{3} (F_{cd} F^{cd} +
\frac{12}{l^2}) \eta_{ab} &=& 0, \cr
d\star F + \frac{2}{\sqrt{3}} F \wedge F &=& 0.
\end{eqnarray}
where our convention for the Hodge dual is $\epsilon_{01234} =1$. 
We will now lift this to a ten-dimensional solution. The lift formula
is \cite{cejm} (see also \cite{cvetic})
\begin{eqnarray}
\label{tendlift}
ds^2_{10} &=&
ds^2_5+ l^2 \sum_{i=1}^3 \left[ (d\mu_i)^2 + \mu_i^2
\left( d\xi_i + \frac{2}{l\sqrt{3}} A \right)^2 \right], \cr
F^{(5)} &=& (1+ *_{(10)}) \left[
-\frac{4}{l} {\rm{vol}_{(5)}} + \frac{{\it{l}}^2}{\sqrt{3}}
\sum_{i=1}^3 d(\mu_i^2) \wedge d\xi_i \wedge *_{(5)} F^{(2)} \right]\,,
\end{eqnarray}
where $\mu_1 = \sin\alpha$, $\mu_2 = \cos\alpha\, \sin\beta$, $\mu_3 =
\cos\alpha\, \cos\beta$ with $0\le \alpha\le \pi/2$, $0\le\beta\le
\pi/2$, $0\le \xi_i\le 2\pi$ and together they parametrise $S^5$. Note
that we define the Hodge star of a $p$-form $\omega$ in $n$-dimensions
as $*_{(n)}\omega_{i_1\dots i_{n-p}}=\frac{1}{p!}\epsilon_{i_1\dots
i_{n-p}}{}^{j_1\dots j_p}\omega_{j_1\dots j_p}$, with
$\epsilon_{0123456789}=1$ and $\epsilon_{01234}=1$ in an orthonormal
frame. The ten-dimensional geometry is specified by
(\ref{eq:nhgframe}) together with  
\begin{eqnarray}
\label{vbs}
e^5 &=& l \, d\alpha,\qquad\qquad
e^6 = l \, \cos\alpha \, d\beta,
\cr
e^7 &=&  l \, \sin\alpha \, \cos\alpha \, [ d\xi_1 - \sin^2\beta \,
d\xi_2 - \cos^2\beta \, d\xi_3 ],
\cr
e^8 &=& l \, \cos\alpha \, \sin\beta \, \cos\beta \, [ \, d\xi_2 -
  d\xi_3 ],
\cr
e^9
&=& -\frac{2}{\sqrt 3}A\, -l\, \sin^2\alpha \, d\xi_1
- l \, \cos^2\alpha \, ( \sin^2\beta \, d\xi_2 + \cos^2\beta \, d\xi_3).
\end{eqnarray}
and the five form \cite{cejm, cvetic, ggs}
\begin{eqnarray}
\label{nhg5form}
F^{(5)} &=&-4 l^{-1} [ e^0 \wedge e^1\wedge e^2 \wedge e^3 \wedge e^4
  + e^5 \wedge e^6 \wedge {e}^7 \wedge {e}^8 \wedge {e}^9 ] \cr 
&& +\frac{2}{\sqrt 3}(e^5\wedge e^7+e^6\wedge
e^8)\wedge(*_{(5)}F^{(2)} -e^9\wedge F^{(2)}) \cr
&=& - \frac{4}{l} (e^{01234} + e^{56789}) - \frac{1}{l} (e^{57} +
  e^{68}) \wedge [ -3 e^{023} + e^{014} - \frac{2}{\omega}
  \lambda ~ e^{234} \cr
&& + e^9 \wedge (3 e^{14} - e^{23} -
  \frac{2}{\omega} \lambda ~ e^{01}) ]\,.
\end{eqnarray}
%
%
%
%
%
\section{The Killing spinor}
In this section, we will solve the Killing spinor equation. The
strategy will be to use the integrability condition to simplify the
equations on a projected subspace. The ten-dimensional Killing spinor
equation is \cite{ggs}
\be
D_m\epsilon+{i\over 1920}\Gamma^{n_1 n_2 n_3 n_4 n_5}\Gamma_m
F^{(5)}_{n_1 n_2 n_3 n_4 n_5}\epsilon=0\,.
\ee
We record the useful identity
\bea
&&\!\!\!\!\!\!\!\! {i\over 1920} \Gamma^{n_1 n_2 n_3 n_4
  n_5}F^{(5)}_{n_1 n_2  n_3 n_4 
  n_5} \Gamma_m={i\over 4 l}[\Gamma_{01234}-{1 \over 4}
  (\Gamma^{57}+\Gamma^{68})(3 
  \Gamma_{023}-\Gamma_{014}-{2\lambda \over
    \omega}\Gamma_{234})](1+\Gamma_{11})\Gamma_m\nonumber \\
&\equiv& {1\over 2}{\cal M} (1+\Gamma_{11}) \Gamma_a e^a_m\,.
\eea
Here $m$ is a spacetime index while $a$ is a tangent-space index.
The integrability condition is \cite{ggs}
\bea
&&[R_{mns_1s_2}-{1\over 48} {F^{(5)}}_{ms_1
    r_1r_2r_3}{F^{(5)}}_{ns_2}^{~~r_1r_2r_3}]\Gamma^{s_1
  s_2}\epsilon\nonumber \\
&+&[{i\over 24} \nabla_{[m}F^{(5)}_{n]s_1s_2s_3 s_4}+{1\over 96}
  F^{(5)}_{mnr_1 r_2 s_1} F^{(5)r_1 r_2}_{~~~~s_2 s_3s_4}]\Gamma^{s_1
  s_2 s_3 s_4}\epsilon=0\,.
\eea
Using a computer algebra program it can be shown that these imposes
the constraints
\bea
i\Gamma^{0149}\epsilon = \Gamma^{2357}\epsilon =
\Gamma^{2368}\epsilon=-\Gamma^{5678}\epsilon=-\epsilon\,,\\ 
\Gamma^{23}\epsilon=\Gamma^{57}\epsilon=\Gamma^{68}\epsilon=-i\epsilon\,. 
\eea
Of these only three are independent which may be chosen to be
\bea 
\Gamma^{0149}\epsilon=i\epsilon\,,\quad
\Gamma^{23}\epsilon=-i\epsilon\,,\quad
\Gamma^{57}\epsilon=-i\epsilon\,.  
\eea
From these projections it follows that the solution in
(\ref{tendlift}, \ref{nhg5form}) preserves at most 4 supersymmetries
of the possible 32.  After some tedious but straightforward algebra,
one can verify that on the constrained subspace the components of the
Killing spinor equation simplify to :
\bea
\left(\partial_t-{4i\lambda r\over \omega^2 l}\Gamma_4\Gamma_0
P_+\right)\epsilon&=& 0\,,\\ 
\left(\partial_r+[-{3\over 2\lambda}{\omega\over 2 r}
  \Gamma_{04}+{1\over 2 r}\Gamma_{09}-{3\over 2 \lambda}{\omega\over
    2r}\Gamma_{49}]\right)\epsilon&=&0\,,\\
\partial_\phi\epsilon= 0\,,\quad
\partial_\theta\epsilon= 0\,,\quad
\partial_\psi\epsilon&=& 0\,,\\
\partial_\alpha\epsilon =0\,,\quad
\partial_\beta\epsilon&=&0\,,\\
\left(\partial_{\xi_j}+{i\over 2}\right)\epsilon&=&0\,,{~~{\rm for~~}
  j=1,2,3}\, ,
\eea
where we define the projectors 
$$
P_\pm={1\over 2}(1\pm \Gamma_{09})\,,
$$
so that $P_+ P_-=0=P_- P_+$. All angular equations can be easily
solved. This leads to the Killing spinor ansatz
\be
\epsilon=e^{-{i\over 2}(\xi_1+\xi_2+\xi_3)}\epsilon(r,t)\,.
\ee
Then the solution to the $t$ equation is
\be
\epsilon(r,t)=e^{{2i\lambda r t\over \omega^2 l} M_1}\epsilon_r(r)\,,
\ee
where
$M_1=-(\Gamma_{49}+\Gamma_{04})=-\Gamma_{49}(1+\Gamma_{09})
=-2\Gamma_{49}P_+$ and satisfies $M_1^2=0$. Plugging 
this into the $r$ equation leads to
\be
\partial_r \epsilon_r={3\omega \over 2 \lambda
  r}\Gamma_{49}P_+\epsilon_r-{1\over 2r}(P_+-P_-)\epsilon_r\,.
\ee
Now writing $\epsilon_r=\epsilon^+_r +\epsilon^-_r$ such that
$\Gamma_{09}\epsilon^\pm=\pm\epsilon^\pm$ we get
\bea
\partial_r \epsilon^+_r&=&-{1\over 2r}\epsilon^+_r\,,\\
\partial_r \epsilon^-_r&=& {3\omega \over 2 \lambda
  r}\Gamma_{49}\epsilon^+_r+{1\over 2r} \epsilon^-_r\,.
\eea
The first of these immediately leads to 
\be
\epsilon^+_r={1\over \sqrt{r}} \epsilon^+_0\,,
\ee
where $\epsilon^+_0$ is a constant positive chirality spinor.
Plugging this into the second equation leads to
\be
\epsilon^-_r=\sqrt{r}\epsilon^-_0-{3\omega \over 2 \lambda
  \sqrt{r}}\Gamma_{49}\epsilon^+_0\,,
\ee
where $\epsilon^-_0$ is a constant negative chirality spinor.  Thus
the complete Killing 
spinor is
\be
\epsilon=e^{-{i\over 2}(\xi_1+\xi_2+\xi_3)}
\left(\sqrt{r}\epsilon^-_0+{1\over \sqrt{r}}(1-{4i\lambda rt\over
  \omega^2 l}\Gamma_{49}P_+)(1-{3\omega\over 2
  \lambda}\Gamma_{49})\epsilon^+_0\right)\,.
\ee
Here $\epsilon_0^\pm$ are subjected to the same projection conditions
as $\epsilon$. The novelty here compared to the full black hole is the
appearance of the other chirality $\epsilon_0^+$ in the solution.
Alternatively, this result can be expressed compactly as:
\be
\epsilon=e^{-{4i\lambda r t\over \omega^2 l}\Gamma_{49}P_+}
e^{({3\omega\over 2 \lambda}\Gamma_{49}P_+-{1\over 2}\Gamma_{09})\ln
  r}\epsilon_0\,.
\ee
It is sometimes useful to split the solution in terms of $\Gamma_{09}$
chiralities: 
\bea
\label{poincareks}
\epsilon^+&=& e^{-{i\over 2}(\xi_1+\xi_2+\xi_3)} {1\over
  \sqrt{r}}\epsilon_0^+\,,\nonumber\\
\epsilon^-&=& e^{-{i\over 2}(\xi_1+\xi_2+\xi_3)}
\left(\sqrt{r}\epsilon_0^--{1\over \sqrt{r}} ({3\omega\over 2
  \lambda}+{4 i \lambda r t\over \omega^2
  l})\Gamma_{49}\epsilon_0^+\right)\,.
\eea
Thus we conclude that the 10-dimensional lift of the near-horizon
geometry of the black hole under consideration preserves precisely
four supersymmetries with the explicit Killing spinors in
eqs.(\ref{poincareks}). We next turn to computing the isometry
superalgebra of this geometry.
%
%
%
%
\section{Isometry supergroup}
In the present section we shall need the basis vectors dual to the
ten-dimensional frame of the near-horizon geometry in Poincar\'e
coordinates. These are:
\bea \tilde{e_0}&=&{\omega\over 2
r}\partial_t-{1\over
l}(\partial_{\xi_1}+\partial_{\xi_2}+\partial_{\xi_3})\,,\quad
\tilde{e_1}= {2\lambda r\over l \omega} \partial_r\,,\\ \tilde{e_2}&=&
{1\over \omega}\left({2\sin\phi}\partial_\theta
+{2\cot\theta\cos\phi}\partial_\phi-{2\cos\phi\,{\rm
cosec~}\theta}\partial_\psi\right)\,,\\ \tilde{e_3}&=&{1\over
\omega}\left({2\cos\phi}\partial_\theta
-{2\cot\theta\sin\phi}\partial_\phi+{2\sin\phi\,{\rm
cosec~}\theta}\partial_\psi\right)\,,\\ \tilde{e_4}&=&{3\omega^2\over
4\lambda r}\partial_t+{2l\over
\omega\lambda}\partial_\phi-{2\omega\over \lambda
l}(\partial_{\xi_1}+\partial_{\xi_2}+\partial_{\xi_3})\,,\quad
\tilde{e_5}={1\over l}\partial_\alpha\,,\quad \tilde{e_6}={1\over
l}\sec\alpha\partial_\beta\,,\\ \tilde{e_7}&=&{1\over
l}\left(\cot\alpha\partial_{\xi_1}
-\tan\alpha\partial_{\xi_2}-\tan\alpha\partial_{\xi_3}\right)\,,\quad
\tilde{e_8}={1\over l}\left(\cot\beta\sec\alpha\partial_{\xi_2}
-\tan\beta\sec\alpha\partial_{\xi_3}\right)\,,\\
\tilde{e_9}&=&-{1\over
l}(\partial_{\xi_1}+\partial_{\xi_2}+\partial_{\xi_3})\,.  \eea
Following the prescription in \cite{gmt1,gmt2}, we now turn to the
computation of the Killing spinor bilinears $\bar \epsilon \Gamma^i
\epsilon$ which are Killing vectors. For the ten-dimensional complex
Weyl representation of definite $\Gamma_{09}$-chirality, one can show
that \be \bar\epsilon^\pm \Gamma^a \epsilon^\pm=0\,, \ee unless $a=0$ or
$a=9$. Conversely \be \bar\epsilon^\mp \Gamma^a \epsilon^\pm=0\,, \ee if
$a=0$ or $a=9$.  Define \be c=-{3\omega\over 2\lambda}-{4i\lambda r
t\over \omega^2 l}\,.  \ee First consider $a=I$ where $I\neq 0$ or
$9$. We have \be \bar \epsilon\Gamma^I \epsilon= \bar\epsilon_0^-
\Gamma^I \epsilon_0^++\bar\epsilon_0^+ \Gamma^I \epsilon_0^-+{1\over
r}\bar\epsilon_0^+ (c\Gamma^I\Gamma_{49}-c^*
\Gamma_{49}\Gamma^I)\epsilon_0^+ \,.  \ee Next consider $a=z$ where
$z=0$ or $9$. This gives \be \bar\epsilon\Gamma^z \epsilon= r
\bar\epsilon_0^- \Gamma^z \epsilon_0^-+ \bar\epsilon_0^-c\Gamma^z
\Gamma_{49}\epsilon_0^+-\bar\epsilon_0^+ c^*
\Gamma_{49}\Gamma^z\epsilon_0^- + {1\over r}\bar\epsilon_0^+
(\Gamma^z-c c^* \Gamma_{49}\Gamma^z \Gamma_{49})\epsilon_0^+\,.  \ee
Thus we have, \bea (\bar\epsilon \Gamma^a \epsilon) \tilde e_a &=&
\bar\epsilon_0^- \Gamma^0 \epsilon_0^- r(\tilde e_0-\tilde e_9) +
\bar\epsilon_0^-\Gamma_4 \epsilon_0^+(c(\tilde e_0-\tilde e_9)+\tilde
e_4+i\tilde e_1)+\bar\epsilon_0^- \Gamma^A \epsilon_0^+ \tilde
e_A\nonumber \\ &+& \bar\epsilon_0^+ \Gamma_4 \epsilon_0^- (c^*(\tilde
e_0-\tilde e_9)+\tilde e_4-i\tilde e_1)+\bar\epsilon_0^+\Gamma^A
\epsilon_0^-\tilde e_A\nonumber \\ &+& \bar\epsilon_0^+\Gamma^0
\epsilon_0^+ {1\over r}(\tilde e_0+\tilde e_9+(c+c^*)\tilde
e_4-i(c-c^*)\tilde e_1+c c^* (\tilde e_0-\tilde e_9)) +
\bar\epsilon_0^+\Gamma_{49}\Gamma^{A}\epsilon_0^+ {1\over
r}(c-c^*)\tilde e_{A}\,,\nonumber \\ \eea
where $A$ takes all values from 2 to 8 except 4. The terms involving
$A$ vanish as we show now. Note that the spinor $\epsilon_0^\pm$ can
be written as $(1+i
\Gamma_{23})(1+i\Gamma_{57})(1+i\Gamma_{68})\epsilon_0/8$ so that we
may always pull a suitable one of these projectors through $\Gamma^A$
which then changes its chirality and annihilates the conjugate on the
left. We have therefore shown that there are four independent
coefficients of the form $\bar \epsilon_0^\pm \Gamma^a \epsilon_0^\pm$
contained in the Killing spinor bilinears. This demonstrates that
there are four bosonic generators in the superisometry group. In other
words, each of these generators corresponds to the coefficient of a
certain linear combination of $z_1 z_1^*$, $z_1 z_{32}^*$, $z_1^*
z_{32}$ and $z_{32}z_{32}^*$ where $z_1,z_{32}$ are taken as the two
independent complex components of the Killing spinor.  Thus the
independent Killing vectors are:
\bea k^{(1)}&=& r (\tilde e_0-\tilde
e_9)\,,\\ k^{(2)} &=& c(\tilde e_0-\tilde e_9)+\tilde e_4+i\tilde
e_1\,,\\ k^{(3)} &=& c^* (\tilde e_0-\tilde e_9)+\tilde e_4-i\tilde
e_1\,,\\ k^{(4)} &=& {1\over r}(\tilde e_0+\tilde e_9+(c+c^*)\tilde
e_4-i(c-c^*)\tilde e_1+c c^* (\tilde e_0-\tilde e_9))\,.  
\eea 
These vectors can be easily verified to be Killing. Expressed in the
holonomic frame these are:
\bea 
k^{(1)} &=& {\omega\over 2}\partial_t\,,\\ k^{(2)} &=&
{-2i\lambda\over \omega 
l} (t\partial_t-r\partial_r)+{2l\over \omega
\lambda}\partial_\phi-{2\omega\over \lambda
l}\partial_{\underline{\xi}}\,,\\ k^{(3)} &=& {2i\lambda\over \omega
l} (t\partial_t-r\partial_r)+{2l\over \omega
\lambda}\partial_\phi-{2\omega\over \lambda
l}\partial_{\underline{\xi}}\,,\\ k^{(4)} &=& {\omega
(3l^2+\lambda^2)\over 8\lambda^2}{1\over r^2}\partial_t+8
{\lambda^2\over \omega^3 l^2} t^2 \partial_t-{16\lambda^2\over
l^2\omega^3} rt\partial_r-{6l\over \lambda^2}{1\over
r}\partial_\phi-{2l\over \lambda^2}{1\over
r}\partial_{\underline{\xi}}\,.  
\eea 
All these Killing vectors are null.
Rescaling $k^{(j)}$ by ${2i\lambda\over \omega l}$ and defining ${1\over
  2}(k^{(2)}-k^{(3)}) = {\cal J}$, ${1\over
  2}(k^{(2)}+k^{(3)}) = Z$, $k^{(1)} = E^+$,
  $k^{(4)}= E^-$,
we get the non-zero commutators
\bea
\left[{\cal J},E^\pm  \right] = \pm\,E^{\pm}\,, & \left[{\cal J},Z
  \right] = 
\left[Z,E^\pm  \right]=0\,, & \left[E^+,E^-  \right] = 2{\cal
  J}\,. 
\eea
This is the algebra $\mathfrak{sl}(2,R)\oplus\mathfrak{u}(1)$ where
$E^\pm, {\cal J}$ are the generators of $\mathfrak{sl}(2,R)$ and $Z$
is the generator of the $\mathfrak{u}(1)$ R-symmetry.  Just as there
is a bosonic charge $Q_B(k)$ associated with each isometry $k$ of the
solution, to each Killing spinor $\epsilon$, there corresponds a
fermionic charge $Q_F(\epsilon)$. The algebra of these is encoded in
the decomposition of the bilinears constructed above in terms of the
bosonic charges \cite{gmt1,gmt2} (see also \cite{achar,fig}). To
extract the decomposition in a convenient form, we define the two
linearly independent Killing spinors
\bea
 \epsilon^{(1)} &=&\frac{1}{\sqrt{r}}e^{-\frac{i}{2}(\xi_1 + \xi_2 +
   \xi_3)}\left[\epsilon_0^+ + c \Gamma_{49}\epsilon_0^+
   \right]\,\\
\epsilon^{(2)} &=& \sqrt{r}e^{-\frac{i}{2}(\xi_1 + \xi_2 +
   \xi_3)}\epsilon_0^-\,.
\eea
We obtain the following linearly independent bilinears
\be
\begin{array}{ll}\label{eq:bilinearse1e2}
\left(\bar\epsilon^{(2)}\Gamma^a\epsilon^{(1)}  \right)\tilde e_a =
(\bar\epsilon_0^-\Gamma^4\epsilon_0^+)\,k^{(2)}\,,\qquad
& \left(\bar\epsilon^{(2)}\Gamma^a\epsilon^{(2)}  \right)\tilde e_a =
(\bar\epsilon_0^-\Gamma^0\epsilon_0^-)\,k^{(1)}\\
\\
\left(\bar\epsilon^{(1)}\Gamma^a\epsilon^{(2)}  \right)\tilde e_a =
(\bar\epsilon_0^+\Gamma^4\epsilon_0^-)\,k^{(3)}\,,\qquad
&\left(\bar\epsilon^{(1)}\Gamma^a\epsilon^{(1)}  \right)\tilde e_a =
(\bar\epsilon_0^+\Gamma^0\epsilon_0^+)\,k^{(4)},. 
\end{array}
\ee
Let us define the fermionic generators associated to the Killing
spinors as follows
\bea
\epsilon^{(1)}&\rightarrow\,\, Q^{(1)} \qquad
  \epsilon^{(2)}&\rightarrow\,\,Q^{(2)}\,. 
\eea
Then it immediately follows from (\ref{eq:bilinearse1e2}) that
\be
\begin{array}{llll}
\left\{ \bar Q^{(2)} ,Q^{(1)} \right\} = Z + {\cal J}\,,&   \left\{ \bar
Q^{(1)} ,Q^{(2)} \right\} = Z - {\cal J}\,,& \left\{ \bar Q^{(2)} ,Q^{(2)}
\right\} = E^+\,,& \left\{ \bar Q^{(1)} ,Q^{(1)} \right\} = E^{-} \,.
\end{array}
\ee
All other odd-odd anti commutators are zero. 
These are in the standard $\mathfrak{sl}(2|1)$ form
\cite{Frappat:1996pb}.
In addition to $k^{(1)},\ldots ,k^{(4)}$, there are also bosonic
isometries of 
this solution which are not associated with the supergroup.
To this end it can also be verified that
the left-invariant vector fields (which generate right translations)
\bea
\xi_1^R&=&
-\sin\psi\partial_\theta -\cot\theta\cos\psi\partial_\psi+\cos\psi{\rm
  cosec~}\theta\partial_\phi \,,\cr
\xi_2^R&=&\cos\psi\partial_\theta-\cot\theta\sin\psi\partial_\psi
+\sin\psi{\rm cosec~}\theta\partial_\phi \,,\cr
\xi_3^R&=& \partial_\psi\,,
\eea
satisfying $[\xi_i^R,\xi_j^R]=-\epsilon_{ijk}\xi_k^R$ are Killing,
reflecting the $\mathfrak{su}(2)_R$ isometries of the squashed $S^3$
of the 
near-horizon region.
In addition to these we expect there to be more bosonic isometries
coming from the $S^5$ part of the geometry that preserve the 1-form
$i\sum_{k=1}^3 \bar z_i \, dz_i$ where $z_i = l \mu_i e^{i\xi_i}$ as
before with $\mu_1 = \sin\alpha$, $\mu_2 = \cos\alpha \sin\beta$ and
$\mu_3 = \cos\alpha \cos\beta$. The following can be seen to be
Killing vectors of our geometry.
\begin{eqnarray}
J_{13}+J_{24} &=& - \cos \xi_{12} [\sin\beta \, \partial_\alpha - 
  \tan\alpha \cos\beta \, \partial_\beta] + \sin \xi_{12} [\cot\alpha
  \sin\beta \, \partial_{\xi_1} + \tan\alpha \csc\beta \,
  \partial_{\xi_2}], \cr 
J_{14}-J_{23} &=& ~~\, \sin\xi_{12} [ \sin\beta \, \partial_\alpha -
  \tan\alpha \cos\beta \, \partial_\beta ] + \cos\xi_{12} [ \cot\alpha
  \sin\beta \, \partial_{\xi_1} + \tan\alpha \csc \beta \,
  \partial_{\xi_2}], \cr
J_{15} + J_{26} &=& - \cos\xi_{13} [ \cos\beta \, \partial_\alpha +
  \tan\alpha \sin\beta \, \partial_\beta] + \sin\xi_{13} [ \cot\alpha
  \cos\beta \, \partial_{\xi_1} + \tan\alpha \sec\beta \,
  \partial_{\xi_3}], \cr
J_{16} - J_{25} &=& ~~\, \sin\xi_{13} [ \cos\beta \, \partial_\alpha + 
  \tan\alpha \sin\beta \, \partial_\beta] + \cos\xi_{13} [ \cot\alpha
 \cos\beta \, \partial_{\xi_1} + \tan\alpha \sec\beta \,
  \partial_{\xi_3}], \cr
J_{35} + J_{46} &=& -\cos\xi_{23} \, \partial_\beta + \sin\xi_{23}
  [\cot\beta \, \partial_{\xi_2} + \tan\beta \, \partial_{\xi_3}], \cr
J_{36} - J_{45} &=& ~~\, \sin\xi_{23} \, \partial_\beta + \cos
  \xi_{23} [\cot\beta \, \partial_{\xi_2} + \tan\beta \,
  \partial_{\xi_3}], \cr 
J_{12} &=& \partial_{\xi_1}, ~~~~ J_{34} = \partial_{\xi_2}, ~~~~ J_{56}
  = \partial_{\xi_3}\,.
\end{eqnarray}
where $\xi_{ij} = \xi_i - \xi_j$. These form the algebra $\mathfrak{u}(3)$. 
The
algebra can be calculated using 
\begin{equation}
[J_{ij}, \, J_{mn}] = \delta_{in} J_{jm} + \delta_{jm} J_{in} -
\delta_{im} J_{jn} - \delta_{jn} J_{im}.
\end{equation}
We have checked that the Lie-derivative of the five form along all the
above Killing vectors vanishes. Thus we have demonstrated that the 
isometry superalgebra of our near-horizon geometry is
$\mathfrak{su}(1,1|1)\oplus \mathfrak{su}(2)\oplus \mathfrak{u}(3)$.  Hence, we conclude that the isometry supergroup is $SU(1,1|1)\times SU(2)\times U(3)$.
%
%
%
%
%
\section{Global coordinates}
We will now consider global coordinates.
Let us first rewrite the five-dimensional part of the metric in
Poincar\'e-like coordinates as follows: 
\begin{equation}
ds^2= - \frac{4 (1+ \frac{3\omega^2}{l^2})}{\omega^2
  (1+\frac{3\omega^2}{4l^2})} r^2 \, dt^2  + \frac{\omega^2}{4
  (1+\frac{3\omega^2}{l^2})} \frac{dr^2}{r^2} + \frac{\omega^2}{4}
  ((\sigma_1^L)^2 + (\sigma_2^L)^2) \\+ \frac{\omega^2}{4}
  (1+\frac{3\omega^2}{4l^2}) [ \sigma_3^L + \frac{6}{\omega l (1+
  \frac{3\omega^2}{4l^2})} r \, dt]^2.  
\end{equation}
We perform the coordinate transformation{\footnote{To cover the full
    range of $r,t$  the range of $\rho$ and $\tau$ should be between
    $-\infty$ to $\infty$.}}
\begin{equation}
t = \frac{\sqrt{b^2 + \rho^2}~\sin \frac{\tau}{b}}{a [- \rho +
  \sqrt{b^2 + \rho^2} ~\cos \frac{\tau}{b}]}, ~~~ r = - \rho +
  \sqrt{b^2 + \rho^2} ~ \cos \frac{\tau}{b}\,,
\end{equation}
\begin{equation}
\tilde \phi := \phi + \frac{6 a b^3}{\omega l} \log \frac{b +
\sqrt{b^2 + \rho^2} ~ \sin \frac{\tau}{b}}{b \cos \frac{\tau}{b} 
- \rho \sin \frac{\tau}{b}}\,.
\end{equation}
Here
$a^2 =\frac{4 \lambda^2}{\omega^2 l^2
(1+\frac{3\omega^2}{4l^2})}$ and $b^2 = \frac{\omega^2 l^2}{4
\lambda^2}$.
This brings the metric into the form
\begin{equation}
ds^2 = - (1 + \frac{\rho^2}{b^2}) \, d\tau^2 + \frac{d\rho^2}{1 +
\frac{\rho^2}{b^2}} + \frac{\omega^2}{4} ((\tilde\sigma_{1}^{L})^2 +
({\tilde\sigma_{2}}^{L})^2)  
+ \frac{\omega^2}{4} (1+\frac{3\omega^2}{4l^2}) (\tilde \sigma_{3}^L -
\frac{6 a b}{\omega l} \rho \, d\tau)^2  \,,
\end{equation}
where $\tilde\sigma^L_i$'s have $\tilde \phi$ in their definition.
The AdS$_2$ part of the metric is now manifestly in global form. The
gauge field reads  
\begin{equation}
A = -\frac{\sqrt{3}}{2} \left[ \frac{\omega^2}{4l} \tilde \sigma_{3}^L -
  \frac{2}{\omega a b} \rho \, d\tau \right]\,,
\end{equation}
after a gauge transformation. We choose the tangent space basis to be
\begin{eqnarray}
e^0 = f~ d\tau, ~~~ e^1 =
f^{-1} d\rho, ~~ e^2 = \frac{\omega}{2}
\tilde\sigma_1^L, ~~ e^3 = \frac{\omega}{2} \tilde\sigma_2^L, ~~~ e^4 =
\frac{\omega}{2ab}  ~ (\tilde\sigma_3^L -
\frac{6ab}{\omega l} \rho \, d\tau)\,.
\end{eqnarray}
where $f=\sqrt{1 +\frac{\rho^2}{b^2}}$. For notational convenience we will drop the tilde from now.
%
%
%
In this basis the field strength and its Hodge dual associated with
$A$ are 
\be
F =- \frac{\sqrt{3}}{2} \left[ \frac{2}{\omega a b} ~ e^{01} -
  \frac{1}{l} e^{23} \right], \quad  
\star F =  \frac{\sqrt{3}}{2} \left[ \frac{2}{\omega ab} ~ e^{234} +
  \frac{1}{l} e^{014} \right]. 
\ee
These satisfy the equation $d \star F + \frac{2}{\sqrt{3}} F
\wedge F = 0$. 
After the 10-dimensional lift, the five-form reads
\be
F^{(5)}=-\frac{4}{l} (e^{01234} + e^{56789}  ) +
\frac{1}{l}\left(e^{57} + e^{68}   \right)\wedge [ e^{014} -
e^{239} + \frac{2 l}{\omega a b}\left(e^{234} + e^{019} \right) ]\,.
\ee
The projection conditions following from integrability in global
coordinates turn out to be: 
\bea
\Gamma^{0149}\epsilon&=&-i\epsilon\,,\quad
\Gamma^{2357}\epsilon=\Gamma^{2368}\epsilon
=\Gamma^{5678}\epsilon=\epsilon\,,\\ 
\Gamma^{23}\epsilon&=&i\epsilon\,,\quad
\Gamma^{57}\epsilon=\Gamma^{68}\epsilon=-i\epsilon\,.
\eea
showing, again, that at most four supersymmetries are preserved by the
near-horizon geometry. Note that these conditions are almost the same
as, but nevertheless different from, the corresponding ones in
Poincar\'e coordinates. The flux contributes 
\bea
&&{i\over 1920} \Gamma^{n_1 n_2 n_3 n_4 n_5}F^{(5)}_{n_1 n_2 n_3 n_4
  n_5} \Gamma_m={i\over 4 l}[\Gamma_{01234}+{1 \over 4}
  (\Gamma^{57}+\Gamma^{68})(- 
  \Gamma_{014} +\frac{2 l}{\omega
    ab}\Gamma_{234})](1+\Gamma_{11})\Gamma_m\nonumber \\ 
&\equiv& {1\over 2}{\cal M}_G (1+\Gamma_{11}) \Gamma_a e^a_m
\eea
to the Killing spinor equation. Here $m$ is a spacetime index while
$a$ is a tangent-space index. 
Using these we get the following simplified component equations:
\bea
\left(\partial_\tau-{i\rho\over 2 b^2}\Gamma_{49}+f (\frac{1}{\omega
  ab}\Gamma_{19} + \frac{3i}{2l}\Gamma_{09})   \right)\epsilon=0\,,\\  
\left(\partial_\rho+\frac{1}{l f}\hat M\right)\epsilon&=&0\,,\\
\partial_\theta\epsilon=0\,,\quad
\partial_\phi\epsilon=0\,,\quad
\partial_\psi\epsilon&=&0\,,\\
\partial_\alpha\epsilon=0\,,\quad
\partial_\beta\epsilon&=&0\,,\\
\left(\partial_{\xi_j}+{i\over 2}\right)\epsilon&=&0\,,{~~{\rm for~~}
  j=1,2,3}\,. 
\eea
where $\hat M = \frac{2b}{l} (\frac{3}{2} \Gamma_{04} +
\frac{l}{\omega ab}  ~ \Gamma_{09})$,
which satisfies $\hat M ^2 = 1$.
Again, all the angular equations are trivial and may be integrated
immediately. Let us now solve the $\rho$ equation to write  
\begin{equation}
\epsilon (\tau, \rho) = e^{- \frac{1}{2} \sinh^{-1} \frac{\rho}{b} ~
  \hat M} \epsilon (\tau)\,,
\end{equation}
 where  $\sinh^{-1} x = \log[x + \sqrt{1+x^2}]$. Then to solve the $\tau$
equation let us first rewrite the equation in the following form
\begin{equation}
\partial_\tau \epsilon = \frac{i}{2b} [ \frac{\rho}{b} \Gamma_{49} -
  f ~ \hat M \Gamma_{49}] \epsilon\,,
\end{equation}
where we make use of the projection $\Gamma_{0149} \epsilon = i
\epsilon$ to eliminate $\Gamma_{19}$ in favour of $\Gamma_{04}$. Then
it is straightforward to verify that the spinor
\begin{equation}
\epsilon (\tau, \rho) = e^{- \frac{1}{2} \sinh^{-1} \frac{\rho}{b} \,
  \hat M } e^{-\frac{i}{2} \hat M \Gamma_{49} \frac{\tau}{b}}
  \epsilon_0 \,,
\end{equation}
where $\epsilon_0$ satisfying all the projections conditions is a
solution to the Killing spinor equation.
This solution can be split in terms of $\hat M\epsilon_0^\pm=\pm
\epsilon_0^\pm$ as 
\be
\epsilon=(e^{-{\chi\over 2}}\cos{\tau\over 2 b}+i e^{\chi\over
  2}\sin{\tau\over 2 b}\Gamma_{49})\epsilon_0^++(e^{{\chi\over
    2}}\cos{\tau\over 2 b}-i e^{-{\chi\over 
  2}}\sin{\tau\over 2 b}\Gamma_{49})\epsilon_0^-\,,
\ee
where $\rho=b\sinh\chi$.
\subsection*{Supergroup in global coordinates}
The supergroup in global coordinates can be computed in the same
manner as was done in the Poincar\'e coordinates.  The basis vectors
dual to the global vielbein are the same as in Poincar\'e coordinates
with the exception of 
\begin{eqnarray}
\tilde{e_0}&=&\frac{1}{f}\partial_\tau + \frac{6 ab}{l\omega
  f}\rho\partial_\phi -     \frac{2 ab}{\omega l f} \rho
(\partial_{\xi_1}+\partial_{\xi_2}+\partial_{\xi_3})\,,\quad 
\tilde{e_1}= f \partial_\rho\,,\\
\tilde{e_4}&=&\frac{2ab}{\omega}\partial_\phi    + \frac{\omega ab}{2
  l^2}(\partial_{\xi_1}+\partial_{\xi_2}+\partial_{\xi_3})\,. 
\end{eqnarray}
In the same way as in section 4, one can use the constraints from the
integrability condition to show that the only nonzero bilinears are
$(\bar\epsilon\Gamma_0\epsilon)$,$(\bar\epsilon\Gamma_1\epsilon)$,
$(\bar\epsilon\Gamma_4\epsilon)$ and
$(\bar\epsilon\Gamma_9\epsilon)$. In addition we can use the condition
$\hat M \epsilon_0^\pm = \pm \epsilon_0^\pm$ to derive the following
relations 
\begin{equation}\label{globalrels}
\begin{array}{ll}
\left(\bar\epsilon_0^\pm\Gamma_9\epsilon_0^\pm\right)
=\frac{2l}{3\omega ab}(\bar\epsilon_0^\pm \Gamma_4
\epsilon_0^\pm)\,,\qquad 
& \left(\bar\epsilon_0^\pm \Gamma_9 \epsilon_0^\mp\right) = -
\frac{3\omega ab}{2l} \left(\bar\epsilon_0^\pm \Gamma_4
\epsilon_0^\mp\right)\\  
\\
\left(\bar\epsilon_0^\pm\Gamma_0 \epsilon_0^\pm  \right) = \mp
\frac{l}{3 b} (\bar\epsilon_0^\pm \Gamma_4 \epsilon_0^\pm)\,,\qquad 
&\left(\bar\epsilon_0^\pm\Gamma_1 \epsilon_0^\mp\right) =
\pm\frac{i  a \omega}{2}(\bar\epsilon_0^\pm \Gamma_4\epsilon_0^\mp)\,. 
\end{array}
\end{equation}
With the aid of these we compute the independent bilinears. \newpage
\begin{eqnarray}
&&(\bar\epsilon \Gamma^a \epsilon)\, \tilde e_a=\nonumber \\
 && (\bar\epsilon_0^+\Gamma_4 \epsilon_0^-) \left[\frac{i  a\omega}{2
 }\cos\frac{\tau}{b} \tilde e_1   - \frac{i
 a\omega}{2}\frac{\rho}{b}\sin\frac{\tau}{b}\tilde e_0 + (\frac{3 i
 \omega ab}{2l}f\sin\frac{\tau}{b} +1 )\tilde e_4 + (\frac{-3\omega
 ab}{2l} + if\sin\frac{\tau}{b})\tilde e_9\right]\nonumber\\ 
&&+(\bar\epsilon_0^- \Gamma_4\epsilon_0^+) \left[\frac{-ia\omega}{2
 }\cos\frac{\tau}{b} \tilde e_1   +
 \frac{ia\omega}{2}\frac{\rho}{b}\sin\frac{\tau}{b}\tilde e_0 +
 (\frac{-3 i \omega ab}{2l}f\sin\frac{\tau}{b} +1 )\tilde e_4 +
 (\frac{-3\omega ab}{2l} - if\sin\frac{\tau}{b})\tilde e_9\right]
 \nonumber\\ 
&&+(\bar\epsilon_0^+\Gamma_4\epsilon_0^+)\left[ -\frac{l}{3
 b}\sin\frac{\tau}{b}\tilde e_1  -\frac{l}{3 b}
 (\frac{\rho}{b}\cos\frac{\tau}{b} - f)\tilde e_0  +
 (f\cos\frac{\tau}{b} - \frac{\rho}{b})\tilde e_4 + \frac{2 l}{3\omega
 ab}(f\cos\frac{\tau}{b}-\frac{\rho}{b}) \tilde e_9
 \right]\nonumber\\ 
&&+(\bar\epsilon_0^-\Gamma_4\epsilon_0^-)\left[ -\frac{l}{3
 b}\sin\frac{\tau}{b}\tilde e_1  -\frac{l}{3 b}
 (\frac{\rho}{b}\cos\frac{\tau}{b} + f)\tilde e_0  +
 (f\cos\frac{\tau}{b} + \frac{\rho}{b})\tilde e_4 + \frac{2 l}{3\omega
 ab}(f\cos\frac{\tau}{b}+\frac{\rho}{b}) \tilde e_9
 \right]\nonumber\,.\\ 
\end{eqnarray} 
 We have checked that these are Killing vectors of the near-horizon metric.  
%
Expressed in the holonomic basis these are:

\begin{eqnarray}
v^{(1)} &=& -\frac{i a \omega}{2 b f} \rho
\sin\frac{\tau}{b}\partial_\tau + \frac{i a \omega f}{2}
\cos\frac{\tau}{b}\partial_\rho + (\frac{2ab}{\omega} +
\frac{3ia^2b^2}{l f}\sin\frac{\tau}{b})\partial_\phi + (\frac{2 ab
  \omega}{l^2} - \frac{i a^2 b^2}{f
  l}\sin\frac{\tau}{b})\partial_{\underline{\xi}} \nonumber\\
v^{(2)}&=& v^{(1)\,*}\nonumber \\
v^{(3)}&=&
\frac{l}{3bf}(f-\frac{\rho}{b}\cos\frac{\tau}{b})\partial_\tau -
\frac{fl}{3b}\sin\frac{\tau}{b}\partial_\rho + \frac{2a b}{\omega
  f}\cos\frac{\tau}{b}\partial_\phi\nonumber -\frac{2ab}{3\omega
  f}\cos\frac{\tau}{b}\partial_{\underline{\xi}}\\ 
v^{(4)} &=&
-\frac{l}{3bf}(f+\frac{\rho}{b}\cos\frac{\tau}{b})\partial_\tau  -
\frac{fl}{3b}\sin\frac{\tau}{b}\partial_\rho + \frac{2a b}{\omega
  f}\cos\frac{\tau}{b}\partial_\phi -\frac{2ab}{3 \omega
  f}\cos\frac{\tau}{b}\partial_{\underline{\xi}}\,. 
\end{eqnarray}
The generators of the purely bosonic isometries do not change in the
global coordinates.

%
%
%
%
%
%
\section{Poincar\'e D-brane probes}
In this section we initiate the study of probe branes in the
near-horizon geometry.  To establish our conventions we quote here the
D3-brane action we shall be working with: 
\be S_{\text{D3}} = -\left[
\int_{\text{D3}} \text{dvol} \pm C^{(4)} \right]\,.
\ee 
In this expression, dvol is the volume form associated to the induced
metric on the 
world volume, which we denote by $h$, and $C^{(4)}$ is the pull
back of the four-form 
potential. The positive sign is for a brane and the negative sign for
an anti-brane.  The conserved charges will be specified using the
point particle Lagrangian denoted by $L$ obtained after integrating
over all the spatial coordinates. From a world-volume
perspective, supersymmetry of a configuration can be established by
studying the kappa-symmetry 
condition. We say that an (anti-) brane is supersymmetric if it obeys
an equation of the form 
\begin{equation}
\Gamma\epsilon = \pm  i \epsilon\,.
\end{equation}
The negative sign is for a brane and the positive sign for an
anti-brane. The spinor $\epsilon$ is the background Killing spinor
derived above. Here $\Gamma$ is the kappa-projection matrix, defined
as 
\begin{eqnarray}\label{eq:kappasymmetryprojector}
\nonumber\Gamma&=&{1\over 4!} {1\over
  \sqrt{-h}}\epsilon^{\sigma_i\sigma_j\sigma_k\sigma_l}
  \gamma_{\sigma_i\sigma_j\sigma_k\sigma_l}\,,\\  
&=& -{1\over \sqrt{-h}}
(\gamma_0\gamma_1\gamma_2\gamma_3+(-h_{01}
  \gamma_{23}-h_{03}\gamma_{12}+h_{02}\gamma_{13}
  +h_{13}\gamma_{02}-h_{12}\gamma_{03}-h_{23}\gamma_{01})\nonumber  
\\ &+&(h_{23}h_{01}+h_{12}h_{03}-h_{13}h_{20}))\,, 
\end{eqnarray}
and $\gamma_{\sigma_i}$ are the world volume gamma matrices
\begin{equation}
\gamma_{\sigma_i}=\partial_{\sigma_i} X^\mu \Gamma_\mu\,.
\end{equation}
and $\gamma_{\sigma_i\sigma_j} =
\frac{1}{2}(\gamma_{\sigma_i}\gamma_{\sigma_j}
-\gamma_{\sigma_j}\gamma_{\sigma_i})$.  
The world-volume gamma matrices satisfy
$\{\gamma_{\sigma_i},\gamma_{\sigma_j}\}=2 
h_{\sigma_i\sigma_j}$.
As in (\ref{eq:kappasymmetryprojector}), we will sometimes find it
convenient to use the shorthand $\gamma_i = \gamma_{\sigma_i}$ for
world-volume indices. 
\subsection{Solving the equations of motion}
In Poincar\'e coordinates one can write the 5-form RR field
strength as $F^{(5)} = dC^{(4)}$ where
\begin{eqnarray}
C^{(4)} &=& \frac{2\omega}{\lambda} e^{0234} + \cot \alpha \,
  e^{678} \wedge (e^9 +
  \frac{2}{\sqrt{3}} A) \cr
&& \!\!\!\!\!\!\!\!\!\!\!\!\!\!\!\!\!\!\!\!\!\!\!\!\! -
  \frac{2}{\sqrt{3}} \left[ A \wedge (e^{57}+e^{68}) \wedge 
(e^9 +  \frac{2}{\sqrt{3}} A) + \frac{l}{2} (e^9 +
  \frac{2}{\sqrt{3}}A) \wedge (\star F +  
  \frac{2}{\sqrt{3}} A \wedge F) \right]\,, 
\end{eqnarray}
with
\begin{equation}
\star F + \frac{2}{\sqrt{3}} A \wedge F = \frac{\sqrt{3}}{l} \left[
  e^0 \wedge (e^{23} - e^{14}) + \frac{l\omega^2}{4} (1 +
  \frac{2\omega^2}{l^2}) \sigma_{123} \right]\,.
\end{equation}

\subsubsection{Giant probes}

Let us now turn to probe D3-branes that wrap a sub-manifold of the
deformed $S^5$ part of the geometry similar to the giant gravitons of
pure AdS. We choose the following static-gauge ansatz 
\begin{equation}\label{eq:giants_poincare}
t = \sigma_0, ~~~ \beta = \sigma_1, ~~~ \xi_2 = \sigma_2, ~~~ \xi_3 =
\sigma_3 
\end{equation}
with the rest of the coordinates assumed to be functions of $\sigma_0$
only. The DBI part of the action follows from\footnote{In this section
  we quote the full action for completeness. There are terms that can
  be dropped consistently from the action without changing the
  equations of motion for the class of solutions we are interested
  in. We will drop such terms from now on to avoid clutter.} 
\begin{eqnarray}
&&\nonumber\sqrt{-\det h_{\sigma_i\sigma_j}}=
  \frac{l^2}{4\omega}\cos\sigma_1 \sin\sigma_1\cos^3\alpha \Biggl[
    \cos^2\alpha \left( \omega^3\Sigma_3 + 8lr \right)^2  - 64 l
    \omega r \left(\omega^2\Sigma_3 - l^2 \sin^2\alpha \dot\xi_1
    \right) \Biggr.  \\ 
&&\Biggl.- 4\omega^2 \left( \omega^2\Sigma_3^2 (l^2 + \omega^2)  +
  \omega^2 l^2 \left( \sin^2\theta \dot\psi^2 + \dot\theta^2 + 2
  \Sigma_3\sin^2\alpha \dot\xi_1 \right)  + 4l^4 \left( \dot\alpha^2 +
  \sin^2\alpha \dot\xi_1^2 \right)\right) - \frac{4l^4\omega^4 \dot
  r^2}{\lambda^2 r^2}\Biggr]^{1/2}\nonumber\\ 
\end{eqnarray}
where $\Sigma_3 = \dot\phi + \cos\theta\dot\psi$. The WZ coupling for
these configurations is 
\begin{equation}
C^{(4)}_{\sigma_0\sigma_1\sigma_2\sigma_3}=l^3 \left(l \dot \xi_1 +
  \frac{2r}{\omega} + 
  \frac{\omega^2}{4l} \Sigma_3\right) \cos^4 \alpha \sin\sigma_1
  \cos\sigma_1 \,.
\end{equation}
It can be verified that for $\dot \xi_1 = \dot\theta=\dot \phi = \dot
\psi = \dot \alpha = \dot r=0$ all 
equations of motion for an anti-brane are satisfied
identically. These giant-like
solutions carry non-zero angular momentum given by
\begin{equation}
P_\phi = 2\pi^2 l^2\omega^2 \cos^2 \alpha, ~~~ P_{\xi_1} = 2\pi^2
l^4 \cos^2 \alpha.\,~~~P_{\psi}=2\pi^2l^2\omega^2\cos^2\alpha \cos\theta\,.
\end{equation}
The giant like solutions found here have $H=0$. Note that $P_{\phi}|_{\rm max}=2\pi^2 l^2\omega^2,~P_{\xi_1}|_{\rm max}=2\pi^2 l^4$, and $P_{\psi}|_{\rm max}=2\pi^2 l^2\omega^2$ which suggest a stringy exclusion principle at work.
\subsubsection{Dual-giant probes}
Now we look for solutions that are analogous to the dual giant
gravitons in AdS in that their world volume takes up an $S^3$ in the
five-dimensional part of our geometry. 
Choosing static gauge, our ansatz is
\bea\label{eq:poincare_dual_giants}
t = \sigma_0, ~~~ \theta = \sigma_1, ~~~ \phi = \sigma_2,
~~~ \psi = \sigma_3
\eea
with all other coordinates assumed to be functions of $\sigma_0$
only. Thus the DBI contribution to the action follows from 
\bea
\sqrt{-\det h_{\sigma_i\sigma_j}} = \frac{\omega^{5/2}}{16 \, l}  \sin
\sigma_1 \left[\omega (8r + l \omega \sum_{i=1}^3 \mu_i^2 \, \dot
  \xi_i)^2 - 4l (l^2 + \omega^2) \sum_{i=1}^3 \mu_i^2 \, \dot \xi_i
  (4r + l\omega \,  \dot \xi_i) \right]^{1/2}\,.
\eea
Without loss of generality we have dropped terms involving
$\dot\alpha, \dot \beta$ and $\dot r$ that do not contribute to the
equations of motion for the configurations we are about to study. 
The pull back of the four-form potential is
\begin{eqnarray}
C^{(4)}_{\sigma_0\sigma_1\sigma_2\sigma_3} = - \left[\frac{4r
\omega^3}{l} + l^2 \omega^2 (1 + \frac{\omega^2}{2l^2})
\sum_{i=1}^3 \mu_i^2 \dot \xi_i \right] \, \frac{1}{8}\sin\sigma_1   
\end{eqnarray}
%
%
%
To find solutions we first note that since the Lagrangian depends only
on $\dot \xi_i$'s putting $\ddot \xi_i =0$ would solve the $\xi_i$ e.o.m
.
%
%
Setting $\dot\xi_i=0$ solves the equations of motion and gives  
the Hamiltonian 
$H = - L$. We find for the momenta conjugate to the
angular variables $P_{\xi_i}=\frac{\partial L}{\partial \dot\xi_i}$
\begin{equation}
P_{\xi_i}= 3 \pi^2 \omega^2 l^2 (1 + \frac{\omega^2}{3l^2})
\mu_i^2 \,.
\end{equation}
This means that $ \sum_{i=1}^3 P_{\xi_i} = 3 \pi^2 \omega^2
l^2 (1 + \frac{\omega^2}{3l^2})$ on our solutions. Furthermore we
find that $H=0$. 

If we use the coordinates $\tilde t$ and $\tilde \phi$, which give the
asymptotic geometry of $AdS_5 \times S^5$ in the standard global
coordinates, then we see that the vanishing Hamiltonian actually
implies $E = -\frac{2}{l} J$ where $J$ is the spin of the probe branes
when measured in the new coordinates. When one considers multiple
configurations of dual-giants in $AdS_5\times S^5$ there is an upper
limit on their number given by the number of units of flux through the
5-sphere\cite{nvs, benasmith}. In our case too one expects that there
is an upper limit on the number of dual-giants.
\subsection{Supersymmetry}
Let us now investigate the kappa symmetry conditions for the
configurations introduced above. 
\subsubsection{Giant Probes}
For the solutions (\ref{eq:giants_poincare}) we find the world-volume
gamma matrices 
\bea
\gamma_0&=& {4r\over \omega}\Gamma_0 P_+\,,\\
\gamma_1 &=& l\cos \alpha\Gamma_6\,,\\
\gamma_2&=& -l\cos\alpha\sin\sigma_1 (-\Gamma_8
\cos\sigma_1+(\Gamma_9\cos\alpha+\Gamma_7\sin\alpha)\sin\sigma_1)\,,\\
\gamma_3&=& -l\cos\alpha\cos\sigma_1(\Gamma_8\sin\sigma_1
+(\Gamma_9\cos\alpha+\Gamma_7\sin\alpha)\cos\sigma_1)\,,
\eea
On the solution $\sqrt{-h}=\frac{1}{\omega}\,2 l^3 r \cos^4
\alpha \sin\beta\cos\beta$. Thus, using equation
(\ref{eq:kappasymmetryprojector}) we get 
\begin{equation}
\Gamma=i \sec\alpha [-2\Gamma_0 P_+ (\cos\alpha \Gamma_9 - \sin\alpha
  \Gamma_7) - \cos\alpha]\,. 
\end{equation}
And hence
\be
\Gamma\epsilon=i\epsilon\,,
\ee
for $\epsilon = P_-\eta$, $\eta$ being the Killing spinor in Poincar\'e
coordinates with $P_+\eta=0$. This sets $\epsilon_0^+=0$. Hence these
configurations are half-BPS
with respect to the near-horizon preserving precisely the
supersymmetries of the full black hole. The isometry preserved by the
brane 
can be determined by adopting a similar procedure as in section
4. The Killing vector preserved by the brane is
proportional to $\partial_t$ which is just the Hamiltonian. Equating
this to zero gives us the $H=0$ condition obtained from the equations
of motion.

\subsubsection{Dual-giant Probes}
For the solutions (\ref{eq:poincare_dual_giants}) we find the
world-volume gamma matrices  
\bea
\gamma_0&=&{4 r\over \omega} \Gamma_0 P_+\,,\\
\gamma_1&=& {\omega\over 2}
(\sin\sigma_2\Gamma_2+\cos\sigma_2\Gamma_3)\,,\\
\gamma_2 &=& -{\omega\over 4l}(3\omega
\Gamma_0-2\lambda\Gamma_4+\omega\Gamma_9)\,,\\
\gamma_3 &=&\cos\sigma_1\gamma_2-{\omega\over 2}\sin\sigma_1
(\cos\sigma_2\Gamma_2-\sin\sigma_2\Gamma_3)\,.
\eea
On the solution $\sqrt{-h}= \frac{\omega^3}{2l}r\sin\sigma_1$. Using
(\ref{eq:kappasymmetryprojector}), we calculate 
\begin{equation}
\Gamma = \frac{i}{2\omega}\left[ \Gamma_0P_+ (3\omega\Gamma_0 -
  2\lambda\Gamma_4 + \omega\Gamma_9) + 2\omega \right]\,. 
\end{equation}
With this we find
$\Gamma\epsilon = -i\epsilon$ for $\epsilon = P_-\eta$ with $\eta$ the
Killing spinor in Poincar\'e coordinates and $P_+\eta=0$, as in the
previous case.  Hence 
these probes are half-BPS with respect to the near-horizon. As in the
previous case, the Killing spinor bilinear implies $H=0$, consistent
with the equations of motion. Thus both the solutions preserve only
the supersymmetries of the full black hole.
%
%
%
%
%
%
\section{Global D-brane probes}
In this section we exhibit some half-BPS D3-brane probes in the near
horizon geometry in global coordinates.

\subsection{Solving the equations of motion}
In global coordinates we can take the 4-form RR potential to be
\begin{eqnarray}
&& C^{(4)} = \frac{4\rho}{l f} e^{0234} +
  \cot \alpha \, e^{678} \wedge (e^9 + \frac{2}{\sqrt{3}} A) \cr 
&& ~~~~~~  - \frac{2}{\sqrt{3}} \left[ A \wedge (e^{57}+e^{68}) \wedge
  (e^9 + \frac{2}{\sqrt{3}} A) + \frac{l}{2} (e^9 +
  \frac{2}{\sqrt{3}}A) \wedge (\star F + \frac{2}{\sqrt{3}} A \wedge
  F) \right], \cr 
&& \star F + \frac{2}{\sqrt{3}} A \wedge F = \frac{\sqrt{3}}{2} \left[  
  \frac{2 a b}{\omega } (1 + \frac{\omega^2}{2l^2}) e^{234} +
  \frac{2}{l} e^{014} + \frac{2 a b \rho}{\omega l f } e^{023} \right].
\end{eqnarray}
%
\subsubsection{Giant Probes}
We now exhibit a two classes of solutions to the DBI action of the
D3-brane probes in global coordinates. We first choose
\begin{equation}
\tau = \sigma_0, ~~ \beta = \sigma_1, ~~ \xi_2 = \sigma_2, ~~ \xi_3 =
\sigma_3
\end{equation}
with the rest of the coordinates functions of $\sigma_0$. 
The DBI contribution to the action follows from
\begin{eqnarray}
\sqrt{-\det h_{\sigma_i\sigma_j}}&=&\frac{1}{8\omega} l^2 \cos^3\alpha
 \sin 2\beta\Biggl[ {64 \rho^2 l^2\over a^2 b^2}\cos^2\alpha +\frac{8
 l\omega \rho}{ab}\left( -8l^2 
 \sin^2\alpha \dot\xi_1 + 2\omega^2 (\cos^2\alpha - 4) \right) \\ 
&+&  \omega^2 \left( 16l^2 - 16l^4 \sin^2\alpha \dot\xi_1^2  + 8 l^2
 \omega^2 \sin^2\alpha \dot\xi_1\dot\phi + \omega^2\dot\phi^2 (-8l^2 -
 4\omega^2 + \omega^2\cos^2\alpha)\right)\Biggr]^{1/2}\,.\nonumber \\
\end{eqnarray}
The WZ coupling is
\begin{equation}
C^{(4)}_{\sigma_0 \sigma_1 \sigma_2 \sigma_3} =  l^4 \cos^4\alpha
\cos\sigma_1 \sin\sigma_1 \left[ \dot \xi_1 - \frac{\omega^2}{4l^2}
  \Sigma_3+ \frac{2\rho}{\omega l ab} 
  \right]\,, 
\end{equation}
where, as before, $\Sigma_3 = \dot\phi + \cos\theta\dot\psi$ and
without loss of generality we have dropped terms involving $\dot\rho,
\dot\alpha,\dot\theta$ and $\dot\psi$ which do not contribute to the
equations of motion. One can verify that 
\begin{equation}
 |\dot \phi| = \frac{2l}{\omega\lambda}, ~~~ |\dot \xi_1|
= \frac{2\omega}{l \lambda} ,~~~\dot\psi=0
\end{equation}
are solutions to the Lagrangian ${\cal L} = - \sqrt{-\det
h_{\sigma_i\sigma_j}} \pm C^{(4)}_{\sigma_0 \sigma_1 \sigma_2
\sigma_3}$ for any constant value of $\alpha,\psi$ and $\theta$,
provided $\rho>-\rho_g$ for branes and $\rho<\rho_g$ for anti-branes
where 
$$
\rho_g={3\omega  ab^2\over 2 l}\,.
$$
One
must take $\dot\phi$ and $\dot\xi_1$ positive for an anti-brane and
negative for a brane. 
The conserved
charges for these solutions are
\begin{equation}
P_\phi = (\frac{2\pi^2}{3} l^4+{4\pi^2 a\rho\omega^2 l^3\over
  \rho\pm\rho_g}) \cos^2 \alpha, ~~~ 
P_{\xi_1} = 2\pi^2 l^4 \cos^2\alpha\,,
\end{equation}
with the above sign for branes and below for anti-branes.
Note that $P_{\phi}$ is infinite at $\rho=\rho_g$ while $P_{\xi_1}$ is
independent of $\rho$.  We will demonstrate later on that supersymmetry dictates $\rho=0$. For this, the maximum value of the momenta are $P_{\phi}|_{\rm max}={2\pi^2 l^4\over 3},~P_{\xi_1}|_{\rm max}=2\pi^2 l^4$ again suggesting a stringy exclusion principle at work.
It is easy to verify that the Lagrangian vanishes and the Hamiltonian
is given by 
\begin{equation}
H = \frac{2l}{\omega \lambda} |P_\phi| + \frac{2\omega}{l\lambda}
|P_{\xi_1}| 
\end{equation}
This is actually the relation expected for BPS objects. 
To see this
one can verify that the following Killing vector of the background 
\begin{equation}
\label{ggbps}
{2\lambda\over 3 \omega}\partial_\tau +{4 l\over
3\omega^2}\partial_\phi+{4\over 3l}\partial_{\xi_1}
\end{equation}
is preserved by the probe brane solutions above for $\rho=0$. 
This can be
seen by considering the bilinears of the supersymmetries
preserved by the probe branes similar to those in section 4 and
5. Then identifying the generators $\partial_\tau$, $\partial_\phi$
and $\partial_{\xi_1}$ with the charges $H$, $P_\phi$ and $P_{\xi_1}$
respectively in eq.(\ref{ggbps}) and equating it to zero results in the
BPS equation. 

There is another class of solutions which have $\dot \psi \ne 0$ as
well. It is easy to verify that
\begin{equation}
\theta = 0, ~~~ \dot \phi = \dot \psi =  
\frac{\eta \, l }{\omega \lambda}, ~~~ \dot \xi_1 = \frac{2 \eta
  \omega}{l\lambda} 
\end{equation}
and
\begin{equation}
\theta = \pi, ~~~ \dot \phi = - \dot \psi =
\frac{\eta l}{\omega \lambda}, ~~~ \dot \xi_1 = 
\frac{2\eta \omega}{l\lambda} 
\end{equation}
 are solutions to the action ${\cal L} = - \sqrt{- \det h_{\sigma_i
\sigma_j}} + \eta \, C^{(4)}$ for $\eta = \pm 1$ whenever
\begin{equation}
\rho \le \rho_g \,.
\end{equation}
The solutions at $\theta = 0$ have
\begin{eqnarray}
P_{\xi_1} = \eta \, 4\pi^2l^4 \, \cos^2 \alpha, ~~~ P_{\phi} = P_\psi
= \eta \, \frac{2\pi^2l^2 (l^2 + 3 \omega^2 \rho/\rho_g)}{3 (1 -
  \rho/\rho_g)} \cos^2\alpha, 
\end{eqnarray}
and those at $\theta = \pi$ have
\begin{equation}
P_{\xi_1} = \eta \, 4\pi^2l^4 \, \cos^2 \alpha, ~~~ P_{\phi} = -
P_{\psi} = \eta \, \frac{2\pi^2l^2 (l^2 + 3 \omega^2 \rho/\rho_g)}{3 (1 -
  \rho/\rho_g)} \cos^2\alpha.
\end{equation}
These configurations have vanishing Lagrangian and therefore their
Hamiltonian is
\begin{equation}
H = \frac{l}{\omega\lambda}(|P_\phi|+|P_\psi|)+\frac{2\omega}{l\lambda}|P_{\xi_1}|=\frac{2l}{\omega \lambda} |P_\phi|  +
\frac{2\omega}{l\lambda} |P_{\xi_1}|.  
\end{equation}

\subsubsection{Dual-giant Probes}
Let us assume the most general ansatz (in static gauge) for a
dual-giant graviton in global coordinates:
\begin{equation}
\tau = \sigma_0,\quad \theta = \sigma_1,\quad\phi = \sigma_2,\quad
\psi = \sigma_3, 
\end{equation}
where all other embedding coordinates are functions of $\sigma_0$. 
The DBI contribution to the action may be written in the form 
\begin{eqnarray}
\sqrt{-\det h_{\sigma_i \sigma_j}} 
 &=& \frac{\omega^{5/2}}{16 l} \sin\sigma_1 \Biggl( \omega \left[
 \frac{8\rho}{ab} + l\omega\sum_{i=1}^3
 \mu_i^2 \dot\xi_i\right]^2 \Biggr.\nonumber\\ 
& & \Biggl. - 4 (l^2 + \omega^2)\left[ \omega \left( -1 +
 l^2\sum_{i=1}^3 \mu_i^2 \dot\xi_i^2\right) + \frac{4 l \rho}{ab}
 \sum_{i=1}^3 \mu_i^2\dot\xi_i\right] 
 \Biggr)^{1/2}\,. 
\end{eqnarray}
where $h_{\sigma_i\sigma_j}$ denotes the induced metric on the world 
volume of the dual giant.  Without loss of generality we have dropped
terms involving 
$\dot\alpha$, $\dot\beta$ and $\dot\rho$ that do not affect the
equations of motion on the configurations we are about to study.
%
%
%
The induced four-form is
\begin{eqnarray}
C^{(4)}_{\sigma_0 \sigma_1 \sigma_2 \sigma_3} = - \frac{\omega^3}{8}
\left[\frac{1}{l \lambda} (l^2 + \omega^2) - \frac{l}{\omega} (1 + 
  \frac{\omega^2}{2l^2}) \sum_{i=1}^3 \mu_i^2 l \, \dot \xi_i +
  \frac{4\rho}{l ab} \right]
\sin\sigma_1 \,,
\end{eqnarray}
where we have chosen to add a constant for convenience which does not
change the equations of motion. Then one can verify that
\begin{equation}
\dot \xi_1 = \dot \xi_1 = \dot \xi_3 = \frac{2\omega}{l\lambda}
\end{equation}
the equations of motion of the action ${\cal L} = - \sqrt{ - \det
  h_{\sigma_i \sigma_j}} + C^{(4)}$ are satisfied when
\begin{equation}
\rho \le \rho_{dg} := \frac{l a b^2}{2 \omega}\,.
\end{equation}
These solutions have the following conserved charges
\begin{equation}
P_{\xi_i} = \frac{\pi^2\omega^2}{8} \left[ \frac{3\omega^4 + 4 l^2
  (\omega^2 
  + \omega\rho l /(2 a b^2))}{l^2 - 4 \omega\rho
  l /(2 a b^2 ) } - (2l^2 + \omega^2)
  \right] \mu_i^2
\end{equation}
for $i = 1$, $2$ and $3$. Notice that these angular momenta diverge as
the radial positions of the dual-giants $\rho$ approaches
$\rho_{dg}$. Furthermore this critical value of the radial coordinate
is different from what the giants see which is $3\omega^2/l^2$ times
$\rho_{dg}$. The Lagrangian evaluated on the configurations again
vanishes and so the Hamiltonian is given by 
\begin{equation}
 H = \frac{2\omega}{l\lambda} (|P_{\xi_1}| + |P_{\xi_2}| +|P_{\xi_3}|) 
\end{equation}
which also diverges at
$\rho=\rho_{dg}$. The same analysis can be repeated for branes with
$\rho_{dg}\rightarrow -\rho_{dg}$ and changing the signs of
$\dot\xi_i$'s.

\subsection{Supersymmetry}
In this section we analyse the amount of supersymmetry preserved by
the probes in global coordinates. 
\subsubsection{Giant Probes}
First consider $\dot\psi=0$.
The pull-back gamma matrices are given by
\bea
\gamma_0&=&f \Gamma_0-({3\rho\over l}\mp
\frac{2}{a\omega})\Gamma_4+(\pm\frac{b}{l}-{2\rho\over \omega
  ab})\Gamma_9\pm \frac{4 b}{l}\sin\alpha
(\Gamma_7\cos\alpha-\Gamma_9\sin\alpha)\,,\\   
\gamma_1&=&l\cos\alpha\Gamma_6\,,\\
\gamma_2&=&-l\cos\alpha\sin\sigma_1
\left(-\cos\sigma_1\Gamma_8+\sin\sigma_1(\cos\alpha\Gamma_9
+\sin\alpha\Gamma_7)\right)\,,\\ 
\gamma_3&=&-l\cos\alpha\cos\sigma_1
\left(\sin\sigma_1\Gamma_8+\cos\sigma_1
(\Gamma_9\cos\alpha+\Gamma_7\sin\alpha)\right)\,,
\eea
where the upper sign is for an anti-brane and the lower sign for a brane.
Using these we get
\be
\Gamma={i l^3 \cos^3\alpha \sin\sigma_1 \cos\sigma_1\over
  \sqrt{-h}}(\pm\frac{6 b}{l}+{4\rho \over \omega ab})\left({1\over
  2}\cos\alpha+\sin\alpha \Gamma_{79} P^{1,2}_+-\cos\alpha
P^{1,2}_+\right)\,, 
\ee
where 
$$ P^{1,2}_+ = \frac{1}{2}\left[1 + \left(\pm\frac{3b}{l} +
  \frac{2\rho}{\omega ab}\right)^{-1} \left( f\, \Gamma_{09} +
  (-\frac{3\rho}{l}\pm \frac{2}{a \omega})\Gamma_{49} \right)  \right]\,,
$$ 
 and can
be shown to be a projector. We further define the orthogonal projectors to be
$$ P^{1,2}_- = \frac{1}{2}\left[1 - \left(\pm\frac{3b}{l} +
  \frac{2\rho}{\omega ab}\right)^{-1} \left( f \Gamma_{09} +
  (-\frac{3\rho}{l}\pm \frac{2}{a \omega})\Gamma_{49} \right)  \right]\,.
$$ 
Hence, if we choose
$\epsilon=P^{1,2}_-\eta$ with $P^{1,2}_+\eta=0$, $\eta$ being the
Killing spinor in global coordinates, then it is easy to see that
$\Gamma 
\epsilon=\pm i\epsilon$ and that the configurations are thus
half-BPS with respect to the near-horizon.  We also see that the
projectors are ill defined at 
$
\rho=\pm\rho_g
$
where the upper sign is for a brane and lower for anti-brane.
These are the same positions where the equations are not solvable for
the corresponding cases. We must further ensure that
$P^{1,2}_+\eta=0$. 
First consider an anti-brane. Write

$$P_+^1={1\over 2}(1+A \Gamma_{09}+B\Gamma_{49})$$ and the Killing spinor as 
\bea
\eta&=&(e^{-{\chi\over 2}}\cos{\tau\over 2 b}+i e^{\chi\over
  2}\sin{\tau\over 2 b}\Gamma_{49})\epsilon_0^++(e^{{\chi\over
    2}}\cos{\tau\over 2 b}-i e^{-{\chi\over 
  2}}\sin{\tau\over 2 b}\Gamma_{49})\epsilon_0^-\nonumber \\
&=&(f_++i
g_+\Gamma_{49})\epsilon_0^++(f_-+ig_-\Gamma_{49})\epsilon_0^-\,,
\eea
with $\rho=b \sinh\chi$.
Some useful relations are
\bea
\Gamma_{09}\epsilon_0^\pm&=&\pm{2b\over l}(-{3\over
  2}\Gamma_{49}+{l\over \omega a b})\epsilon_0^\pm\,, \\ 
\Gamma_{04}\epsilon_0^\pm&=&\pm {2 b\over l}({3\over 2}+{l\over \omega
  a b}\Gamma_{49})\epsilon_0^\pm\,. 
\eea
We demand $P_+^1\eta=0$  corresponding to $P_-^1\eta$ being preserved.
Now we note that $\epsilon_0^\pm$ and $\Gamma_{49}\epsilon_0^\mp$ have the
same chirality. This leads to the following equations on equating the
coefficient of $\cos{\tau\over 2b}$ 
\bea
e^{-\chi/2}(1+{2A\over \omega
a})\epsilon_0^+&=&-e^{\chi/2}({3bA\over l}+B)\Gamma_{49}\epsilon_0^-\,,\\
e^{\chi/2}({3bA\over l}+B)\epsilon_0^+&=&-e^{-\chi/2}(1+{2A\over \omega
a})\Gamma_{49}\epsilon_0^-\,.
\eea
These lead to the conclusion that $\rho=0$ and
$\epsilon_0^+=-\Gamma_{49}\epsilon_0^-$. 
It can be verified that these conditions satisfy the equations
obtained 
from the coefficients of $\sin{\tau\over 2b}$ as well. The same
calculation can be repeated for the brane case. The conclusion is that 
the condition on the constant spinors is
$\epsilon_0^+=\pm\Gamma_{49}\epsilon_0^-$, the upper sign for brane
and lower for anti-brane and $\rho=0$.


The calculation for non-zero $\dot\psi$ can be repeated in a similar
manner. It turns out that the world-volume gamma matrices are
identical to the above case and hence the supersymmetry analysis is
identical to the one given there. 
\subsubsection{Dual Giants}
The world-volume gamma matrices are
\begin{eqnarray}
\gamma_0 &=& -\frac{2}{l}\rho \left( \frac{3}{2}\Gamma_4 +
\frac{l}{\omega ab}\Gamma_9
\right) + f~\Gamma_0
\pm \frac{4 b}{l}\Gamma_9\\ 
\gamma_1 &=& \frac{\omega}{2}\left(  \cos\phi\Gamma_3 +
\sin\phi\Gamma_2 \right)\\ 
\gamma_2 &=& \frac{\omega}{2l }\left( \frac{\omega}{2}\Gamma_9 +
\frac{l}{ab}\Gamma_4 \right)\\ 
\gamma_3 &=& \cos\theta\gamma_2 + \frac{\omega}{2}\sin\theta\left(
\sin\phi\Gamma_3 - \cos\phi\Gamma_2 \right) \,,
\end{eqnarray}
where the upper sign is for brane and lower for anti-brane.
After some algebra one can derive the following simple expression for
the Kappa-symmetry projection matrix 
\begin{equation}
\Gamma = \frac{i}{\sqrt{-h}}\left(
\frac{\omega}{2}\right)^2\,\sin\theta (h_{02} - \gamma_{0}\gamma_2)\,, 
\end{equation}
with
\begin{equation}\label{dgamma}
h_{02} - \gamma_0\gamma_2 = \frac{\omega}{2 l}\left[ \left(
  \frac{2l}{\omega}\rho   \mp \frac{4}{a}
  \right)\Gamma_{49}  + f~\left(
  \frac{\omega}{2}\Gamma_{09} + \frac{l}{ab}\Gamma_{04} \right)\right] 
\end{equation}
We note that
\begin{equation}
-\det h \mathbb{1}= \left( \frac{\omega}{2} \right)^4 \sin^2\theta
 \left( h_{02} \mathbb{1}- \gamma_0 \gamma_2 \right)^2 = {\omega^6
 \sin^2\sigma_1\over  4 l^2 a^2 b^2}(\rho\pm \rho_{dg})^2 \mathbb{1}
 \,, 
\end{equation}
the upper sign is for brane and lower for anti-brane.
We can thus form the projectors
${\cal P}_\pm = \frac{1}{2}(1 \pm i \Gamma)$. 
From the
above we see that ${\cal P}_\pm$ commutes with $\Gamma^{0149}$,
$\Gamma^{23}$ and $\Gamma^{57}$. 
Furthermore, the projectors become ill defined at $\rho=\rho_{dg}$
which is the same point where the angular momenta blow up.  
The condition on the constant spinors are derived as follows:
For branes, we want to preserve $\Gamma \epsilon=-i\epsilon$.
Let us write 
$\Gamma=c (A\Gamma_{49}+B\Gamma_{09}+C\Gamma_{04})$ where
$c={i\omega^3\over 8 l\sqrt{-h}}\sin\sigma_1$. 
Then after some tedious algebra we get
\bea
{\Gamma\over c}\epsilon&=& \left(f_+(A  -{3b B\over l}+{2 C\over
  \omega a})-ig_+(  {2B\over \omega a}+ {3 b C\over
  l})\right)\Gamma_{49}\epsilon_0^+\nonumber\\ 
&+& \left(-f_-({2 B \over \omega a}+{ 3 b C\over l})-i g_-(A+{3 B
  b\over l}+{2 C\over \omega a})\right)\epsilon_0^-\nonumber \\ 
&+& \left(f_+(2{B\over \omega a}+{3 b C\over l})-i  g_+(A+{3 B b\over
  l}-{2 C\over \omega a})\right)\epsilon_0^+\nonumber \\ 
&+& \left(f_-(A+{3 b B\over l}-{2 C\over \omega a})+i g_-({2 B\over
  \omega a}+{3 b C\over l})\right)\Gamma_{49}\epsilon_0^-\,. 
\eea
Now we equate this to $-{i\over c}\epsilon$. Equating the
$\cos{\tau\over 2 b}$ piece  

\bea
f_+(A-{3 b B\over l}+{2 C\over \omega a})\Gamma_{49}\epsilon_0^+&=&
f_-({2 B\over \omega a}+{3 b C\over l}-{i\over c})\epsilon_0^-\,,\\ 
f_+({2 B\over \omega a}+{3 b C\over l}+{i\over
  c})\epsilon_0^+&=&-f_-(A+{3 b B\over l}-{2C\over\omega a}
)\Gamma_{49}\epsilon_0^-\,. 
\eea
We can read off $A,B,C$ from equation (\ref{dgamma}).
This tells us that $\rho=0$ and
$\epsilon_0^+=\Gamma_{49}\epsilon_0^-$.  
One can check that the other conditions arising from $\sin{\tau\over
  2b}$ are also satisfied. Thus we conclude that, as for the giant
case, supersymmetric dual giants also satisfy $\rho=0$ and
$\epsilon_0^+=\pm\Gamma_{49}\epsilon_0^-$, the upper sign for branes
and lower for anti-branes. 
\subsubsection*{Conserved Killing vector}
The calculation of the Killing vector that the giant and dual giants
preserve is now straightforward. Imposing  
$\epsilon_0^+=\pm\Gamma_{49}\epsilon_0^-$, we get the Killing spinor
to simplify to 
\be
\epsilon=e^{-{i\tau\over 2 b}}(1\pm\Gamma_{49})\epsilon_0^+\,.
\ee
Using this we find
\be
\bar\epsilon\Gamma^a\epsilon\tilde
e_a=2\bar\epsilon_0^+\left(-\Gamma_0\tilde e_0\mp\Gamma_9\tilde e_4\pm
\Gamma_4\tilde e_9\right)\epsilon_0\,. 
\ee
Now using (\ref{globalrels}), the Killing vector becomes
\be
v=-{2\lambda\over 3\omega} \tilde e_0\mp{2 l\over 3 \omega a b}\tilde
e_4\pm\tilde e_9\,, 
\ee
which leads to
\be
H={2\omega\over
  l\lambda}|P_{\underline{\xi}}|+{2l\over\omega\lambda}|P_\phi|\,. 
\ee
This is the expected BPS relation for both giants and dual giants.
%
%
%
%
%
\section{Conclusion}

In this paper, we considered the near-horizon geometry of the simplest
of the supersymmetric $AdS_5$ black holes with two equal angular
momenta and a single U(1) electric charge. It was shown that the
isometry supergroup of the IIB uplift of this black hole is
$SU(1,1|1)\times SU(2)\times U(3)$. This was achieved by explicitly
constructing the Killing spinors of the geometry and then considering
the bilinears following \cite{gmt2}. The near-horizon geometry has
a deformed 3-sphere $\tilde S^3$
and a deformed 5-sphere $\tilde S^5$ with a
fibration of the time coordinate of $AdS_2$ over them. We considered both
Poincar\'e and global-like coordinates for the $AdS_2$ part of the
geometry. We found that the number of supersymmetries of the near-horizon geometry of these black holes is twice that of the full
solution.

We then exhibited several D3-brane configurations in this geometry
that are analogous to the giant and dual-giant gravitons of the $AdS_5
\times S^5$ background. The dual-giant like D3-branes wrap the
deformed-$S^3$ and the giant like objects wrap an $S^3$ inside the
deformed-$S^5$ part of the geometry. In the Poincar\'e coordinates the
branes do not rotate. They still carry non-zero angular momenta. In
global coordinates the branes have to rotate in order to satisfy the
equations of motion. All the configurations considered in Poincar\'e
coordinates preserve two of the four supersymmetries. These two
supersymmetries are simply those of the full black hole solution
restricted to the near-horizon geometry.

We showed that the probes in global coordinates preserve two of the four
supersymmetries of the background when placed at the centre $\rho = 0$
of $AdS_2$ and so are half-BPS. However, the configurations at a
generic non-zero $\rho$ do not preserve any supersymmetries. The
D3-brane probes at generic $\rho$ exhibit interesting features. In
particular, they all satisfy a BPS-like energy condition and see a
critical value of the radial position where their angular momenta
diverge. It will be interesting to understand the physics behind this
behaviour. We expect  there to be more giant-type probe branes like those in \cite{mikhailov}. There is a duality between configurations of giants and dual-giants in $AdS_5\times S^5$. It will be interesting to see if such a duality holds in this case as well.

The results of this paper should help in counting microstates of the
black hole under consideration as mentioned in the introduction. To make further progress in this
direction one has to classify all the BPS objects in the global
coordinates with a given set of supersymmetries. Then one should be
able to quantise them using methods similar to \cite{bglm, ms} (see
also \cite{ast, bfhh, masp, bm}) and count the different
configurations with fixed quantum numbers.

There are several generalisations of the black holes considered here
\cite{cclp1, cclp2, kl} which have non-equal angular momenta in
$AdS_5$ directions and non-equal R-charges in $S^5$ directions (with
one condition among them). However, we suspect that their near-horizon
geometries again preserve  four supersymmetries. The reason is that the
generators of the bosonic part of the isometry group which are
responsible for the generalisation do not participate in the
supersymmetric part of the full supergroup of isometries. We expect
that the near-horizons of the generalisations have the same supergroup
part $SU(1,1|1)$ of the isometries but with the bosonic parts $SU(2)$
and $U(3)$ broken to some subgroups of them. It will be interesting to establish this in detail.

Following Strominger et al \cite{gssy, gsy} one can ask what is the
holographically dual conformal quantum mechanics of the string theory
in the near-horizon geometry of the Gutowski-Reall black holes
considered here. Our superisometries should be an important input in
constructing the Lagrangian for such a quantum mechanics. One also expects that there are
some small black holes with more supersymmetries than the
Gutowski-Reall black holes (see \cite{nvs} for instance). Counting the
microstates in the near-horizon geometry of the Gutowski-Reall black
holes might capture the entropies of the small black holes as well as
in \cite{ss2} in an analogous context. We hope to return to some of
these questions in the future.

%
%
%
%
\section*{Acknowledgements}
We thank Jaume Gomis, Jan Gutowksi, James Lucietti, Hari Kunduri,
Harvey Reall and Claude Warnick for discussions on the subject of this
paper. AS is supported by PPARC and Gonville and Caius college,
Cambridge. JS gratefully acknowledges financial support from the Gates
Cambridge Trust and PPARC. AS thanks the KITP, Santa Barbara, for
hospitality during the final stages of preparation of this
paper. Research at Perimeter Institute is supported in part by the
Government of Canada through NSERC and by the Province of Ontario
through MEDT. This research was supported in part by the National
Science Foundation under Grant No. PHY99-07949.


\end{document}